# BEP Enhancement for Semi-Femtocell MIMO Systems Employing SC-QICs and OSTBCs

Ardavan Rahimian, Farhad Mehran
School of Electronic, Electrical and Computer Engineering
University of Birmingham
Edgbaston, Birmingham B15 2TT, UK

*Abstract*—In mobile cellular networks, it is estimated that more than 60% of voice and data services occur indoors. Therefore, cellular network operators have shown an unprecedented interest in research on femtocell systems from various aspects to extend the indoor wireless coverage for providing high-quality and high data-rate wireless multimedia services contents. In an effort for reducing the bit-error probabilities (BEPs) and also increasing bit/symbol capacity of bandwidth limited error-prone wireless channels in femtocell propagation areas, this paper presents the performance of a promising candidate technology designed based on the state-of-the-art techniques. The performance of powerful space-time turbo codes (STTCs) based on serial concatenation of quadratic interleaved codes (SC-QICs) with the optimal and also suboptimal decoding algorithms, in conjunction with orthogonal space-time block codes (OSTBCs) have been presented in this contribution for wireless multiple-input single-output (MISO), and multiple-input multiple-output (MIMO) semi-femtocells.

*Index Terms*—cellular radio network; diversity; femtocell, forward error correction; multiple-antenna; space-time coding.

## I. INTRODUCTION

Femtocell networks bring significant advantages for both the mobile cellular network operators and the end users. They not only provide indoor coverage for places where macrocell cannot, but they are also capable of offloading traffic from its layer and improve the capacity while significantly save the user equipment's power [1]. However, for supporting the demands for high data-rate video, voice, and data services, it is difficult to achieve simultaneously high transmission quality and high information date-rate in a bandwidth-limited wireless channels. In response to these ever-increasing demands, multiple-antenna systems and technologies based on space-time codes [2-4] have been the most important technological breakthrough in the last two decades [5], which shown to result in substantially higher capacity than their single-antenna counterparts as it is pointed out by Foschini [6]. They have shown to result in very low bit-error probabilities (BEPs) in highly-faded wireless channels, while increasing the data-rates significantly. To provide more coding gains for future mobile systems, the information and coding community also proposed to concatenate the space-time codes with high performance channel codes. In [7-10], space-time codes in cooperation with turbo codes [11] which are the first class of channel codes that perform within one decibels of Shannon were proposed. Hence, these researches motivated us to employ the integrated space-time codes and also turbo codes for improving the physical layer (PHY) performance of generic femtocell wireless communication systems [12], while making further potential enhancements through the following methods: (1) improving the BEP performance in error floor region by using serial concatenation (SC) instead of conventional parallel concatenation schemes; (2) improving the BEP performance in both the waterfall and the floor regions by using permutation algorithm based on the property of quadratic congruence.

## II. CODED M-QAM MIMO SYSTEM MODEL: ANALYSIS

A wireless MIMO system is considered where transmitter and receiver are equipped with *n* and *m* antennas respectively as show in fig. 1. Data are first encoded by turbo encoder, and afterwards, modulated by the multilevel quadrature amplitude (M-QAM) signaling block. Then, the modulated symbols are mapped using space-time block encoder, and transmitted over the channel. At the receiver, the combiner combines received signals that are then sent to the detector. The detected symbols are demodulated at the M-QAM demodulator signaling unit and, are then decoded at the iterative near maximum likelihood (ML) decoder in order to recover the transmitted signal bits.

### A. Coded M-QAM System Model: Construction of SC-QICs

The advent of turbo codes [11] that facilitate the operation of communications systems near Shannon limit was perhaps of the most crucial historic breakthrough after Shannon's seminal work. Since turbo code's invention, they have reached maturity within just two decades due to their interesting near-capacity performance. The initial results showed that the turbo codes could achieve energy efficiencies within only a half decibel of the Shannon capacity. This result was taken into account as an extraordinary performance and first it was met with skepticism, but further research efforts of turbo coding community not only clarified their outstanding performance and characteristics, but also introduced a wealth of practical industrial solutions for wireless communication systems and services [13, 14]. In this work, the focus has been on the turbo codes based on serial concatenation [15] as they have shown to yield performance comparable to the classic parallel code schemes; they are also referred to as serial concatenation of interleaved codes [12, 16].

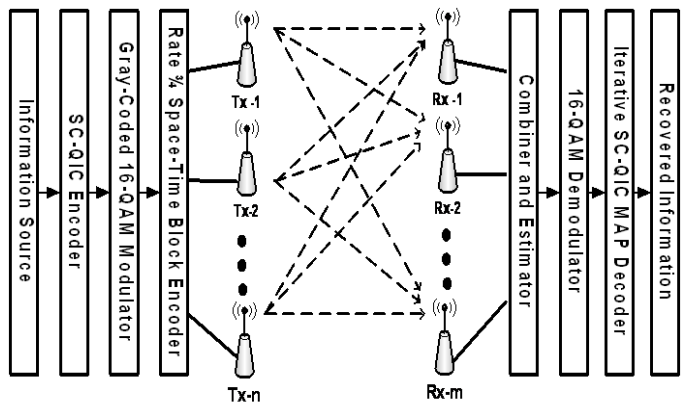

Fig. 1. System Block Diagram for Space-Time SC-QIC Coded M-ary QAM MIMO Wireless Communication System.

As it is noted in [17], the most critical part in the design of turbo codes is the interleaver design. The design of permutation pattern for interleaver has a key role in turbo coded system's exceptional BEP output since substantial increase in minimum distance of the code could be attained from the state-of-the-art design of permutation algorithm. While the interleaver design for the classic turbo codes based on parallel concatenation has attracted a great attention, the interleaver design for serial turbo codes remains scarce. This may be due to the fact they employ concatenation of the different constituent codes with differing distance properties when this algorithm targets the specific utilization for serial scheme [18]. In this work, we will utilize the quadratic interleavers proposed in [19]. They do not only have a straightforward system implementation due to their simple representation in compact form but also result in great performance compared to the matrix-based block interleavers which are of the most crucial commonly used interleaver type in communication systems. In this case, in order to express the permutation for the interleaver of period $T$, and also denoting the permutation vector by $\pi(.)$, take $T = 2^\vartheta$ and choose two constants $0 < \varphi < T$, $k$ odd, and $0 \leq h < T$ [19], then for each $0 \leq \aleph < T$ define cycle vector of $C(\aleph) = mod(\varphi.\aleph.(\aleph+1)/2, T)$, then, $\pi\big(mod(C(N-1)-h,T)\big) = C(0)$ and also for $0 < \aleph < T$, $\pi\big(mod(C(\aleph-1)-h,T)\big) = C(\aleph)$ [20, 21]. Since the mapping to generate the permutation of a sequence of bits is based on the property of quadratic congruence over an integers ring modulo powers of two, the serial turbo codes utilizing this algorithm is referred to as the serial concatenation of quadratic interleaved codes (SC-QICs) as it has been employed in semi-femtocell.

*B. Coded M-QAM System Model: Space-Time Block Coding*

Multiple-antennas based on the space-time codes have been intensively studied since they constitute efficient robust trade-off in terms of their effective throughput, BEP performance, and estimated complexity. After introduction of Alamouti's twin-antenna space-time codes which appealed in terms of the BEP performance and complexity, research was dedicated for developing this scheme for the higher number of transmitters to increase diversity order. Alamouti's scheme was generalized to arbitrary number of transmitter antennas, leading to the notion of STBCs [3, 13]. In this case, for a system with $n$ and $m$ Tx and Rx antennas, radio signals $c_t^i, i = 1, 2, ..., n$, are transmitted simultaneously using $n$ transmit antennas. Assuming the path gain from the transmit antenna $i$ to receive antenna $j$ ($\alpha_{i,j}$) is constant over a frame of length $l$, at time $t$ the radio signal $r_t^j$ received at the antenna $j$ is given by equation (1) below as:

$$\xrightarrow{yields} r_t^j = \sum_{i=1}^{n} \alpha_{i,j}.c_t^i + \gamma_t^i \qquad (1)$$

where $\gamma_t^i$ are the independent samples of a zero-mean complex Gaussian random variable with variance 1/(2SNR) per complex dimension. A STBC is defined transmission matrix $G$ in which its entries are linear combinations of the $x_1, x_2, ..., x_k$ and their conjugates [4]. For Alamouti's scheme and Tarokh et al.'s rate ¾ scheme with three transmitting antennas, the transmission matrices are given as the equations (2) and (3) below [2,4]:

$$G_{2-Tx}^{Alamouti} = \begin{pmatrix} x_1 & x_2 \\ -x_2^* & x_1^* \end{pmatrix} \qquad (2)$$

$$G_{3-Tx}^{rate\ 3/4} = \begin{pmatrix} x_1 & x_2 & \frac{x_3}{\sqrt{2}} \\ -x_2^* & x_1^* & \frac{x_3}{\sqrt{2}} \\ \frac{x_3^*}{\sqrt{2}} & \frac{x_3^*}{\sqrt{2}} & \frac{(-x_1-x_1^*+x_2-x_2^*)}{2} \\ \frac{x_3^*}{\sqrt{2}} & -\frac{x_3^*}{\sqrt{2}} & \frac{(x_2+x_2^*+x_1-x_1^*)}{2} \end{pmatrix} \qquad (3)$$

At the receiver, for detecting symbols of the two transmit antennas (denoted by $s_1$ and $s_2$), the decision metrics given by the equation (4) and the equation (5) have been used as [4]:

$$\left| \left[ \sum_{j=1}^{m} \left( r_1^j \alpha_{1,j}^* + (r_2^j)^* \alpha_{2,j} \right) \right] - s_1 \right|^2 + \left( -1 + \sum_{j=1}^{m} \sum_{i=1}^{2} |\alpha_{i,j}|^2 \right) |s_1|^2 \qquad (4)$$

$$\left| \left[ \sum_{j=1}^{m} \left( r_1^j \alpha_{2,j}^* + (r_2^j)^* \alpha_{1,j} \right) \right] - s_2 \right|^2 + \left( -1 + \sum_{j=1}^{m} \sum_{i=1}^{2} |\alpha_{i,j}|^2 \right) |s_2|^2 \qquad (5)$$

For detecting symbols of the three transmit antennas with rate ¾ (denoted by $s_1$, $s_2$, and $s_3$), the maximum likelihood (ML) decoding of STBCs accounts in order to minimize the decision metrics given by the equations (6), (7), and (8) [4]:

$$\left| \left[ \sum_{j=1}^{m} (r_i^j \alpha_{1,j}^*) + (r_2^j)^* \alpha_{2,j} + \frac{(r_4^j - r_3^j)\alpha_{3,j}^*}{2} \right] - \frac{(r_3^j + r_4^j)^* \alpha_{3,j}}{2} - s_1 \right|^2 + \left( -1 + \sum_{j=1}^{m} \sum_{i=1}^{3} |\alpha_{i,j}|^2 \right) |s_1|^2 \qquad (6)$$

$$\left| \left[ \sum_{j=1}^{m} (r_i^j \alpha_{2,j}^*) - (r_2^j)^* \alpha_{1,j} + \frac{(r_4^j + r_3^j)\alpha_{3,j}^*}{2} \right] + \frac{(-r_3^j + r_4^j)^* \alpha_{3,j}}{2} - s_2 \right|^2 + \left( -1 + \sum_{j=1}^{m} \sum_{i=1}^{3} |\alpha_{i,j}|^2 \right) |s_2|^2 \qquad (7)$$

$$\left| \left[ \sum_{j=1}^{m} \frac{(r_1^j + r_2^j)\alpha_{3,j}^*}{\sqrt{2}} (r_1^j \alpha_{2,j}^*) + \frac{(r_3^j)^*(\alpha_{1,j} + \alpha_{2,j})}{\sqrt{2}} + \frac{(r_4^j)^*(\alpha_{1,j} - \alpha_{2,j})}{\sqrt{2}} \right] - s_3 \right|^2 + \left( -1 + \sum_{j=1}^{m} \sum_{i=1}^{3} |\alpha_{i,j}|^2 \right) |s_3|^2 \qquad (8)$$

International Journal of Electronics Communication and Computer Technology (IJECCT)
Volume 3 Issue 1 (January 2013)

III. MIMO System Model: Simulation and Discussion

Based on the above mentioned discussions, the simulations in terms of BEP versus the energy per bit/noise power spectral density ($E_b/N_0$) are shown for the systems of interest wherein SC-QICs of length 1024 bits in conjunction with Alamouti's twin-antenna space-time codes and also Tarokh et al.'s rate ¾ OSTBCs for three transmit antennas are presented. In this case, inner and outer SC-QIC's constituent codes utilize recursive convolutional codes with generator polynomial metrics $(7,5)_8$ and $(7,5,0; 0,7,5)_8$ with six iterations at the system decoder. The investigation at the system design stage has been carried out for the stochastic channel model. Here, the spectral broadening of the received Rx signal has been modeled based on the Bell spectrum proposed in IEEE 802.11 TGn model given by $S_{(f)} = 1/(1 + A(f/f_d)^2)$ where $A$ is a constant used in order to define $0.1S_{(f)}$ at $f_d$ and also being the Doppler spread ($|f| \leq f_d$) [22]. Meanwhile, the investigation has been based on the worst case scenario wherein the line-of-sight (LoS) path is blocked; hence, Rayleigh distribution is used for modeling the statistical time varying nature of the Rx received signal's envelope. In addition, additive noise at the receivers considered to follow the additive Gaussian noise (AWGN). Although assumptions for channel modeling may seem to be not very precise to some extent, it should also be mentioned that using location-specific models at the radio system design stages is not indispensable since stochastic channel modeling is still the most used tool in order to compare the different techniques and algorithms at the system design and comparison stage. The $M = 16$-QAM digital modulation scheme has been employed as it provides a good trade-off in terms of the power and bandwidth efficiencies; and therefore widely adopted in standardized wireless systems such as in digital video broadcasting (DVB), high speed downlink packet access (HSDPA) protocol, and IEEE 80211a PHY layer for modulating OFDM subcarriers. In this case, instead of binary code mapping, gray code mapping has been used which shown to yield additional coding gains. In the fig. 2, the BEP curves for the discussed systems of interest are presented. In order to indicate the resultant coding gains due to deploying additional antenna with the Tarokh's OSTBCs, we employed one more antenna not only at the transmitter side but also at the receiver side yielding diversity order of six, as compared with that of Alamouti's MISO scheme with order two. As it can be seen from the figure, while the MISO wireless system results in BEP $\approx 10^{-4}$ for SNR = 20 dB, and also employing additional antenna at the transmitter and the receiver based on the Tarokh scheme yields BEP $\approx 10^{-7}$ at the same SNR. The observed difference comes from higher system diversity order which is the main feature of the orthogonal STBCs, and results in the rapid MISO and MIMO wireless STTC systems adaptation.

As it can be seen from the curves, the coding gains due to the utilization of quadratic interleaver is around 4 dB compared to the optimal matrix-based block interleavers with the same size and under the same simulation parameters. This should be considered as an achievement, since they also yield the more undemanding approach in implementation, and also allow the possibility of analysis due to the structured construction. Since in other works it has been shown the optimal result of matrix-based block interleavers is attainable when the depth and span of are relatively high and comparable, it is predicted that the obtained system coding gains are more astonishing when one compares the designed system with the quadratic interleaver, compared with the same system employing rectangular matrix-based block interleavers. In addition, the performance curves for MISO and MIMO systems are presented for the decoding based on original maximum a posteriori (MAP) algorithm, and its simplified version (Max-Log-MAP algorithm) wherein the computational saving arise by making approximation $log(e^{x_1} + \ldots + e^{x_n}) = max\{x_1, \ldots, x_n\}$ in the forward, backward, and state transition metrics computation for evaluating the a posteriori probability (APP) of information bits. As it can be seen, the performance difference between the MAP decoded and Max-Log-MAP decoded systems is not significant. Therefore, these obtained simulation results confirm and also suggest that since numerical stability problems can arise in computations decision metrics in the MAP system decoding due to the large amount of computational complexities, the Max-Log-MAP decoding is a good overall choice when one considers both the error-rate performance and complexity. Since the delay of turbo codes is the main bottleneck of their utilization in many modern delay-sensitive wireless systems and applications, utilization of the Max-Log-MAP is a promising method while keeping the BEP performance penalties reasonable for the intended systems.

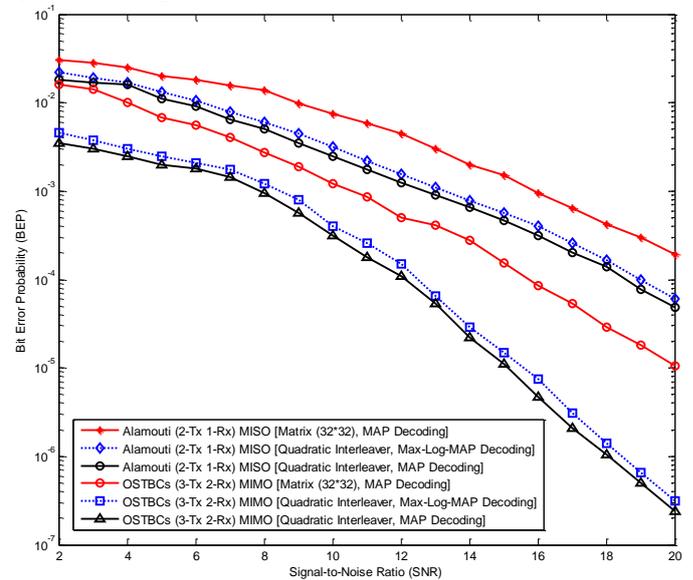

Fig. 2. Bit-Error Probability (BEP) vs. Energy Per Bit/Noise Power Spectral Density ($E_b/N_0$) for Differing Number of Transmitter and Receiver Antennas With $N = 2048$, MAP Max-Log-MAP Algorithms With Six Iterations.

IV. Conclusion

In this contribution, performance of powerful space-time code/channel code combination with outstanding coding gains has been thoroughly evaluated numerically in terms of the BEP for the potential practical utilization in femtocell systems. The following observations can be made from the obtained results. Firstly, it has been revealed that employing additional antenna at the transmitter and the receiver results in outstanding coding gains with realistic decoding complexities proposed by Tarokh et al. Additionally, it has been shown that interleaving based on the property of quadratic congruence over a rings of integers yields outstanding coding gains compared to the matrix-based block interleavers which are the most used interleaver types. The results of this paper can be extended in a number of ways including the investigation on the performance of combined space-time codes based on the LDPC codes for the femtocell and also urban microcell [23] areas. The proposed wireless MISO and MIMO systems can further be optimized using the evolutionary computation optimization techniques as in [24].